cond-mat/9301001

\documentstyle[12pt]{article}

\begin{document}
\begin{center}
{\bf Effect of Loops on the Vibrational Spectrum \\of Percolation Network}

\bigskip
\bigskip

Hisao Nakanishi$^{\dag}$\\
HLRZ, KFA--J\"{u}lich, Postfach 1913\\
W-5170 J\"{u}lich, Germany

\end{center}

\bigskip

\begin{flushleft}
$^{\dag}$Present and permanent address: Department of Physics,
Purdue University,\\West Lafayette, Indiana 47907 U.S.A.
\end{flushleft}

\bigskip

We study the effects of adding loops to a critical percolation cluster
on the diffusional, and equivalently, (scalar) elastic properties of
the fractal network.  From the numerical calculations of the eigenspectrum
of the transition probability matrix, we find that the
spectral dimension $d_s$ and the walk dimension $d_w$ change suddenly
as soon as the floppy ends of a critical percolation cluster are connected
together to form relatively large loops, and that the additional inclusion
of successively smaller loops only change these exponents little if at all.
This suggests that there is a new universality class associated with
the loop-enhanced percolation problem.


\bigskip
\bigskip


\newpage
{\bf I. Introduction}

\bigskip

One of the few experimentally accessible measures of the degree to which
a structure is fractal \cite{mandelbrot,feder} is the spectral dimension
$d_s$ \cite{havlin}.  The spectral dimension was originally defined
\cite{alexander} from the low energy behavior of the vibrational density
of states $n(E)$ of a fractal, elastic network:
\begin{equation}
n(E) \sim E^{d_s /2 -1} \;,
\end{equation}
where $E$ is the energy of a mode.  The energy is proportional to
${\omega}^2$ where $\omega$ is the angular frequency of the vibrational
mode, and writing the above asymptotic relation in terms of the density
of states per unit interval in $\omega$, we have the relation
\begin{equation}
\rho (\omega ) \sim {\omega}^{d_s -1} \; \label{fracton},
\end{equation}
in which $d_s$ has replaced the Euclidean dimension $d$ in a more
familiar equation for the phonon density of states.

The exponent $d_s$ can be measured by various optical and
neutron scattering experiments \cite{courtens},
and for some silica aerogels in particular, a value
in the neighborhood $1.3\pm 0.01$ was found.  Now, such a
fractal is sometimes compared to the critical percolation cluster
\cite{stauffer} since the latter is the simplest and best studied
model of an equilibrium random fractal.  If this comparison
is made for the silica aerogel, the experimental value of $d_s$ happens
to be close to the $d_s$ of the scalar elastic problem \cite{alexander}
on the critical percolation cluster in $d=3$,
and much larger than the corresponding vector elastic result
of about $d_s =0.9$ \cite{mandelbrot}.  (Vector elasticity is the
problem where the displacement of each node in the network is
a {\em vector} rather than a scalar.)  Courtens \cite{courtens}
argued that this is not an evidence for scalar elasticity but
rather that an aerogel has very few {\em floppy endings} unlike
a percolation cluster.  That is, most floppy endings which may
form during the growth of an aerogel are connected together to
form loops during the aging and (supercritical) drying processes.
Thus the final structure is much more rigid with higher value
of $d_s$ although the {\em fractal} dimension $d_f$ is
essentially the same.

In this paper, we test this suggestion, but for the simpler,
scalar elasticity problem, on two- and three-dimensional critical
percolation clusters.  In this case, each bond connecting two
(occupied) nearest neighbor sites of the cluster acts {\em like}
an ideal spring of uniform spring constant in the sense that
it exerts a Hooke's law force based on the difference in the
({\em scalar}) displacements of the two sites.  The effects of
loops are studied by adding controlled amount (and size) of loops
and calculating the dynamic exponents $d_s$ and $d_w$ \cite{havlin}.
As far as we know, this is the first quantitative study of
the effects of adding loops on the structural properties of
a stochastic fractal.

We study this problem as a diffusion problem using
the vibration--diffusion analogy \cite{alexander} for scalar elasticity.
According to this correspondence, e.g., the vibrational spectrum
of the network can be calculated as the spectrum of the
{\em transition probability matrix} ${\bf W}$ \cite{harris,sonali},
where the elements $W_{ij}$ of ${\bf W}$ is simply the hopping
probability per step from node $j$ to node $i$ in the discrete
network.  Thus this matrix ${\bf W}$ contains the information on
both the network topology and the dynamics of the model of
diffusion.  We choose for our calculations the so-called
{\em blind ant} random walk \cite{havlin} as a model of diffusion
although the specific choice of the type of random walk is
irrelevant for our results.  For a blind ant, the random walker
attempts to move without the knowledge of which neighbors are part of
the cluster (thus available) and which ones are not, and thus
when it happens to choose an unavailable neighbor, it cannot
hop and must wait at its current position for the next time step.

{}From the point of view of the diffusion problem, the interesting
quantities include the probability $P(t)$ for a random walker
to return to its starting point after $t$ steps and the root-mean-square
distance $R(t)$ traveled by the random walk in time $t$.
These quantities are expected to have the power-laws for asymptotically
long times $t$:
\begin{eqnarray}
P(t) & \sim & t^{-d_s /2} \\
R(t) & \sim & t^{1/d_w} .
\end{eqnarray}
These relations may be considered to define the exponents $d_s$
and $d_w$.

We calculate these exponents by approximately diagonalizing
the matrix ${\bf W}$ using the method of Ref. \cite{fuchs,sonali}.
Once the diagonalization is done, we compute two
quantities, the density of eigenvalues $n(\lambda )$
(where $\lambda$ denotes the eigenvalues of ${\bf W}$) and a certain
function $\pi (\lambda )$ (which is the product of $n(\lambda )$ and
some coefficient determined when the stationary
initial state distribution is expanded in terms of
the eigenvectors of ${\bf W}$ \cite{jacobs}).
These functions are expected to behave, asymptotically near
$\lambda =1$ \cite{sonali}, as
\begin{eqnarray}
n(\lambda ) & \sim & | \ln \lambda |^{d_s /2-1}
\label{den}  \\
\pi (\lambda ) & \sim & | \ln \lambda |^{1-2/d_w} .
\label{pi}
\end{eqnarray}

\bigskip

{\bf II. Loop Addition}

\bigskip

In this section, we characterize the clusters for which the
calculations described in Section I are performed.
The procedure is as follows:
we first generate random realizations of occupied sites with the
known critical probability $p_c$ (independently of each other) on
either a square or simple-cubic grid of a predetermined size
(edge length $L$).
The percolation problem is then constructed by imposing
the nearest-neighbor connectivity among the occupied sites using
periodic boundary condition in all directions.  We then choose only
those realizations whose maximal cluster spans the grid in all
directions {\em as well as} wraps in all directions, i.e., that
those on which there is a path that winds around the grid (with
the periodic boundaries) in each coordinate direction.
The maximal, wrapping cluster generated in this way is an allowable
starting configuration.

In order to add loops in a controlled manner, we first mark
all perimeter sites which are neighbors to two or more occupied
sites (i.e., those empty sites which would connect two or more
occupied sites if they were occupied).  For each of these
{\em multiple-perimeter sites}, we must decide the size of the
loop it closes.  While it is very easy to determine the size of
the {\em smallest} loop, that is not a very interesting quantity,
since it may also close a much larger loop and such a larger
loop may dominate the rigidity of the structure.  On the other
hand, the size of the {\em largest} loop itself is also not an
interesting quantity since in principle there are many loops
that are embedded in the interior of the cluster with little
influence on the cluster's rigidity.

Thus we use the following
procedure to determine the {\em effective}
size of the loop a multiple-perimeter site closes:
For each such site, we consider every pair of occupied neighbors
{\em in turn}.  For each pair, we calculate the minimum loop
size by a {\em burning} method \cite{herrmann}, spreading a
{\em fire} from each end and determining the first contact of
the two fires.  Then we compute the {\em maximum} among these
minimum loops sizes attached to the perimeter site in question.
This is the appropriate loop size $P$,
the size of the {\em floppiest} loop associated with the particular
multiple-perimeter site.  We then add all multiple-perimeter
sites with $P$ greater than or equal to a certain predetermined
value $P_o$ to control how {\em floppy} a loop must be for it
to be closed. For example, $P_o =\infty$ will close {\em no} loop, and
$P=4$ will close {\em all} loops for both the square and simple cubic
lattices.

In this paper, we compare the results
for $P_o =\infty$, $12$, $8$, and $6$. A typical starting cluster
and the corresponding loop-added clusters on the square lattice
are shown in Fig.1.
Clearly, the fuzzy and floppy endings present in the original
cluster (a) become thicker and better connected successively
in (b), (c), and (d) as more loops are added, giving the impression
of {\em image sharpening}.  The final structure in (d) with $P_o =6$
appears to be a much more solid object than the original;
yet, the fractal dimension of all these clusters are virtually
the same, being about $1.9$.

The above procedure is followed after the starting configuration
is fixed.  The applicable multiple-perimeter sites are ordered
in a particular way (in a typewriter fashion due to the construction
of the cluster), and the addition
procedure applied only {\em once} in the fixed order, with {\em no}
recursion.  That is, once a site is determined to be either added
or not, we do not later consider it again for addition and also
we do not consider perimeter sites newly created by this process
itself.  This removes the problem of the complete filling of
lakes and fjords which would occur particularly when $P_o$ is small,
but does not solve the problem of the final structure depending
somewhat on the initial ordering of the perimeter sites.  Physically,
{\em no recursion} rule might correspond to the assumption that
the formation of additional bonds during aging and drying
is largely simultaneous and {\em not} sequential.

\bigskip

{\bf III. Numerical Results}

\bigskip

Our numerical results for $n(\lambda )$ for the density of
states and $\pi (\lambda )$ (related to the velocity autocorrelation
function of the random walk by a Laplace transform \cite{fuchs,sonali})
are plotted in Fig.2 and Fig.3 respectively.
We see that generally $d=3$ data appear to be behaving
in a clearer way than those of $d=2$.
We summarize the exponent estimates for $d_s$ and $d_w$
in Tables~I and II where the error estimates are mostly from
the least squares regression and do not take into account
any finite size effects or
other systematic errors that may be present.

The grid sizes for the calculations shown in the figures are
$L=100$ for the square lattice and $L=30$ for the simple cubic lattice
(although we have also checked for the cluster size dependence using
smaller grids of $L=50$ for the square lattice and $L=20$ for the
simple cubic lattice).  The spanning and wrapping percolation
clusters (at $p=0.593\approx p_c$ for the square and $p=0.312$
for the simple cubic lattice) contained about $4600$ sites for
$d=2$ and $4000$ sites for $d=3$ before adding any loops.
For $P_o =12$, the size increases to about $5050$ and $4400$,
for $P_o =8$, to about $5250$ and $4900$, and for $P_o =6$,
to $6100$ and $5700$ for $d=2$ and $d=3$, respectively.
For each value of $P_o$, averages were taken over $400$ independent
cluster realizations for the square lattice, and for the simple
cubic lattice, the number of clusters used varied slightly
as $481$, $480$, and $479$ for $P_o =12$, $8$, and $6$, respectively.

Thus, both the size of the clusters and the number of independent
samples in the disorder ensemble are rather modest; however,
they are comparable to the previous calculations by the same method
for the $P_o =\infty$ case \cite{sonali} where the results for $p_c$
were in excellent agreement with other calculations of $d_s$ and $d_w$
as well as with the scaling relation $d_s =2d_f /d_w$
\cite{havlin} (but see also \cite{nakanishi}).
The scope of the calculations were CPU time limited
mainly because of the need to calculate the eigenvalues and
eigenvectors very accurately (usually to 6 digits) for the largest
$200$ or so eigenvalues.  These calculations took typically
3 hours of CPU time on one Cray Y-MP processor
for each $P_o$ for the square
lattice and 6 hours for each $P_o$ for the simple cubic lattice.
Performing these calculations on workstations for long periods of
time would solve the CPU limitation, but then the memory becomes
the limiting factor.

Since the slope in the log-log plot of Fig.2 must equal
$d_s /2-1$ (cf. Eq.(\ref{den})),
it is obvious that the original percolation clusters
($P_o =\infty$) yield completely different values of $d_s$ in
both $d=2$ and $3$ even from the case of $P_o =12$.  Indeed,
the estimate of $d_s$ is about $1.30$ for both $d=2$ and $d=3$
if $P_o =\infty$, but for the loop-added cases, the estimates are
$d_s \approx 1.7$ for $d=2$ and $\approx 2.0$ for $d=3$.  Since the
case $P_o =12$ contains only about $10$ \% more sites and visually
looks very similar to the original percolation cluster
(cf. Fig.1 (a) and (b)), this is a clear
evidence that the structural properties are very sensitive to
the addition of even a few loops.  On the other hand, the slopes
of the data points for the three finite values of $P_o$ do not differ
very much from each other.  Thus once the larger loops are added,
any further addition of smaller loops do not appear to affect the
results very much.

There does seem to be a tendency for
the slope to get steeper when more loops are added.
This tendency is within the statistical standard deviations
for $d=2$ but definitely outside reasonable statistical errors for $d=3$.
A study of smaller clusters ($L=50$ for the square and
$L=20$ for the simple cubic lattices) also shows different
behaviors for $d=2$ and $3$.  For the square lattice,
this slope gets consistently steeper for all $P_o =12$, $8$, and $6$
as $L$ increases; however, the increase in the slope is
well within the statistical standard deviation in all cases.
For the simple cubic lattice, on the other hand,
the slope becomes less steep for $P_o =12$ and steeper for
$P_o =6$ as $L$ increases, the changes being slightly outside
of the statistical errors.  These considerations suggest
that the different $d_s$ for different $P_o <\infty$ might
be a real effect for $d=3$ but less likely to be so for $d=2$.
However, a more complete study, e.g., of the finite size effects
is needed to definitively answer this question.  In any case,
the differences in the slope among the $P_o =12$, $8$, and  $6$
are far smaller than the difference between these values of
$P_o$ and $P_o =\infty$.

Compared to the results for the density of states $n(\lambda )$,
the results for $\pi (\lambda )$ in Fig.3 show a similar,
but much less drastic relative differences in $d_w$ between
the loop-added clusters and the original percolation clusters.
That is, the exponent $d_w$ changes from about $2.9$ for the
original cluster to about $2.4$ for the loop-added case for
$d=2$, a change of only about 17 \% compared to the case of
$d_s$ where the change is more than 30 \%.
This can be understood since the addition of a few
loops may not change the velocity autocorrelation of a random walker
in $t$ steps as much as its chances to return to the starting
point, or put another way, the structural rigidity is much more
sensitive to the addition of a few loops than the overall
random walk displacement on the same network.

Again, the three-dimensional results appear to be better
behaving than the two-dimensional ones.  However, in both
cases, the slopes in the log-log plot of Fig.3 are significantly
steeper for $P_o =\infty$ than for the loop-added clusters, with
a hint of a slight trend for the larger slope for $P_o =6$
in $d=3$.  Since this slope must equal $1-2/d_w$ (cf. Eq.(\ref{pi})),
a larger slope implies a larger value of $d_w$.
These differences among $P_o <\infty$ are, however,
at least as ill-defined as for the case of $n(\lambda )$.

\newpage
\bigskip

{\bf IV. Summary}

\bigskip

In summary, we have presented the analysis of the dynamical
properties of the random walk confined to loop-enhanced
critical percolation clusters.  Through the vibration--diffusion
analogy \cite{alexander}, this work has implications for the
(scalar) elasticity problem on the same fractal network.
{}From the numerical calculations of the eigenspectrum
of the transition probability matrix, we find what appears
to be the crossover to a new universality class
characterized by a significantly
larger spectral dimension $d_s$ and significantly smaller walk
dimension $d_w$ as soon as the floppy ends of a critical
percolation cluster are connected together to form relatively large loops.
The additional inclusion of successively smaller loops does not seem to
further change the exponents $d_s$ and $d_w$ very much.
This result supports the observation of Ref.\cite{courtens} that
the deviation between the experimental observation
of $d_s$ of an aerogel and the previous theoretical calculations
for the {\em vector} elasticity of the critical percolation cluster
is due to the lack of floppy endings in the aerogel
because of the chemical reactions during aging and
(supercritical) drying.

The loop-enhanced percolation problem is reminiscent of
the problem of the external surface
of the critical percolation cluster.  The latter can be defined in
several ways, e.g., by the percolation {\em hull} \cite{stauffer}
and by the perimeter sites accessible to a random walk from the
outside.  The accessibility in that case depends on the size of the
walker \cite{grossman}; in a somewhat similar way,
the {\em floppiness} of the loops
in our problem depends on the minimum loop size $P_o$.  As in our case,
the external perimeter problem also leads to {\em one} new
universality class (in addition to the one for the hull)
in two dimensions \cite{grossman}.
One difference is, however, that our problem
leads to a new universality class also in three dimensions while
this is not so for the external perimeter problem.

Beyond the qualitative analogy, there might lurk a more
quantitative one, since the problem of the external surface is
that of successively controlling the accessibility for a
particle diffusing from outside while our problem is that of
successively controlling the reach of a particle diffusing
on the cluster itself.  This is, however, beyond the scope of
this work and must await further research.

\newpage
\bigskip

{\bf Acknowledgments}

\bigskip

I am grateful to Hans Herrmann for suggesting this problem and
to him and Dietrich Stauffer for helpful discussions.
This work was carried out while I was visiting the HLRZ
Supercomputer Center at KFA--J\"{u}lich in Germany.  I would like to
thank the Center and Hans Herrmann for warm hospitality.

\newpage

\begin{center}
{\bf FIGURES}
\end{center}

\bigskip
\bigskip

\begin{description}
\item[Fig. 1:]
The result of adding successively smaller loops to the critical
percolation cluster is illustrated with:
(a) a typical spanning and wrapping critical percolation cluster on
the square lattice; (b) when all loops of loop parameter $P \ge 12$
are added; (c) when all loops of $P \ge 8$ are added; (d) when all
loops of $P \ge 6$ are added.  The grid size is $100 \times 100$.

\bigskip

\item[Fig. 2:]
Density of eigenvalues $n(\lambda )$ for (a) the square lattice
and (b) the simple cubic lattice.  The symbols $\times$, $\bigcirc$,
$\Box$, and $\Delta$ correspond to $P_o =\infty$, $12$, $8$, and
$6$, respectively.  Lines drawn are linear least squares fits to
the corresponding data shown here. The data for $P_o =\infty$ are
from Ref.\cite{sonali}.

\bigskip

\item[Fig. 3:]
The function $\pi (\lambda )$ for (a) the square lattice
and (b) the simple cubic lattice.  The symbols $\times$, $\bigcirc$,
$\Box$, and $\Delta$ correspond to $P_o =\infty$, $12$, $8$, and
$6$, respectively.  Lines drawn are linear least squares fits to
the corresponding data shown here, and the data for $P_o =\infty$
are from Ref.\cite{sonali}

\end{description}

\newpage

\begin{center}
{\bf TABLES}
\end{center}

\bigskip

\begin{table}[h]
\caption{Numerical estimates of the exponents $d_s$ and $d_w$
for loop-enhanced critical percolation cluster on the square lattice.
The results for $P_o =\infty$ are from Ref.8.
The error estimates are mainly from the least squares regression.}

\bigskip

\begin{tabular}{ccc} \hline\hline
    $P_o$ & $d_s$ & $d_w$ \\
\hline
    $\infty$ & $1.30 \pm 0.02$ & $2.86 \pm 0.04$\\
    $12$ & $1.69 \pm 0.12$ & $2.40 \pm 0.12$\\
    $8$ & $1.66 \pm 0.10$ & $2.42 \pm 0.12$\\
    $6$ & $1.63 \pm 0.11$ & $2.39 \pm 0.15$\\
\hline\hline
\end{tabular}
\end{table}

\bigskip

\begin{table}[h]
\caption{Numerical estimates of the exponents $d_s$ and $d_w$
for loop-enhanced critical percolation cluster
on the simple cubic lattice.
The results for $P_o =\infty$ are from Ref.8.
The error estimates are mainly from the least squares regression.}

\bigskip

\begin{tabular}{ccc} \hline\hline
    $P_o$ & $d_s$ & $d_w$ \\
\hline
    $\infty$ & $1.30 \pm 0.02$ & $3.70 \pm 0.07$\\
    $12$ & $1.99 \pm 0.02$ & $2.67 \pm 0.07$\\
    $8$ & $1.94 \pm 0.02$ & $2.76 \pm 0.08$\\
    $6$ & $1.89 \pm 0.01$ & $2.76 \pm 0.04$\\
\hline\hline
\end{tabular}
\end{table}

\end{document}